\documentstyle[aps]{revtex}

\begin{document}

\draft
\title{Combinatorial Computation of Clebsch-Gordan Coefficients}
\author{Klaus Schertler and Markus H. Thoma}
\address{Institut f\"ur Theoretische Physik, Universit\"at Giessen,
35392 Giessen, Germany}
\date{\today}

\maketitle

\begin{abstract}
The addition of angular momenta can be reduced to elementary
coupling processes of spin-$\frac{1}{2}$-particles. In this way,
a method is developed which allows for a non-recursive, simultaneous
computation of all Clebsch-Gordan coefficients concerning the addition of two
angular momenta. The relevant equations can be interpreted easily,
analogously to simple probabilistic considerations. They provide
an improved understanding of the addition of angular momenta as well as
a practicable evaluation of Clebsch-Gordan coefficients in an easier way
than within the well-known methods.
\end{abstract}


\section{Introduction}


\subsection{Definition and properties of Clebsch-Gordan coefficients}

A system of two quantum mechanical angular momenta is characterized by the
angular momentum operators of the total angular momentum
$\hat{\bf J}^{2}$, $\hat J_{z}$ and the operators of the individual
angular momenta $\hat{\bf J}_{1}^{2}$, $\hat J_{1z}$, $\hat{\bf J}_{2}^{2}$,
$\hat J_{2z}$. These operators can be classified into two groups in which
the corresponding operators commute \cite{saku}:
	\begin{itemize}
                \item  $ \hat{\bf J}_{1}^{2}, \hat{\bf J}_{2}^{2},
                \hat{\bf J}^{2}, \hat J_{z}$,

                \item  $ \hat{\bf J}_{1}^{2}, \hat{\bf J}_{2}^{2},
                \hat J_{1z}, \hat J_{2z} $.
	\end{itemize}
Hence the basis states can be chosen as
	\begin{itemize}
                \item  coupled eigenstates, denoted by $\vert J_{1},J_{2};
                J,M\rangle $ or $\vert J,M \rangle $,
                \item  uncoupled eigenstates, denoted by $\vert J_{1},J_{2};
                M_{1},M_{2}\rangle $ or $\vert (M_{1},M_{2}) \rangle $.
	\end{itemize}
These complete basis systems are related to each other via an unitary
transformation:
	\begin{equation}
                \vert J,M \rangle=\sum_{M_{1},M_{2}} \vert (M_{1},M_{2})
                \rangle
                \langle (M_{1},M_{2}) \vert J,M \rangle .
		\label{trans}
	\end{equation}
The amplitudes $\langle (M_{1},M_{2}) \vert J,M \rangle$ in (\ref{trans})
are called {\em Clebsch-Gordan (CG) coefficients}. General properties
of the CG-coefficients are given e.g. in Ref.\cite{saku}.
Here only a few are listed:
	\begin{itemize}
                \item  CG-coefficients vanish if $M \not= M_{1}+M_{2}$.

                \item  The total angular momentum $J$ satisfies
                $\vert J_{1}-J_{2} \vert \le J \le J_{1}+J_{2}$.

                \item  CG-coefficients can be chosen as real.
        \end{itemize}
%

\subsection{Conventional methods}

A standard method for calculating CG-coefficients  (see e.g. \cite{mess})
is based on the iterative application of the operator
$\hat J_{-} \equiv \hat J_{x}-i J_{y}$ on the maximum state
	\begin{displaymath}
        \vert J, M \rangle = \vert (M_{1} = J_{1},M_{2} = J_{2}) \rangle ,
	\end{displaymath}
where $J = J_{1}+J_{2}$ and $M = M_{1}+M_{2}$. The effect of the
lowering operator on $\vert J, M \rangle$ is given by
	\begin{displaymath}
        \hat J_{-} \vert J, M \rangle = \sqrt{(J+M)(J-M+1)} \vert J, M-1
        \rangle .
	\end{displaymath}
The CG-coefficients are obtained by projecting the resulting states onto
the uncoupled product state $\langle (M_{1},M_{2})\vert$.

Furthermore, Racah gave an elaborate but explicit formula for the
CG-coefficients (see e.g. \cite{mess}). Both the methods are very formal
and render the evaluation even in the case of "small" angular momenta
cumbersome. Thus CG-coefficients have been listed in tables (see e.g.
\cite{cgtables}).

Here we propose an intuitive way allowing for an
immediate computation of CG-coefficients. First we will consider the
special case of coupling of two angular momenta to their maximum angular
momentum, before we will turn to the general case in section III.
Based on these investigations we will provide a very simple way of
calculating CG-coefficients in section IV. In the appendix we will
prove the equivalence of our method with the Racah formula.


\section{Addition of two angular momenta to their maximum angular momentum}

By speaking of a spin-$J$-particle in the following
we denote any object with angular momentum
$J$, e.g. also orbital angular momenta, in order to avoid the
inconvenient expression ``objects with angular momentum $J$''.

\subsection{Decomposition of a particle into spin-$\frac{1}{2}$-particles}
	\label{abschnitt_zerlegen}

According to Schwinger's oscillator model of angular momentum
\cite{saku,schw} every
spin-$J$-particle can be considered as a composition of $2J$
spin-$\frac{1}{2}$-particles.
Then a particle described by the state $\vert J, M \rangle$
with angular momentum $J$ and $z$-component $M$ consists of
	\begin{center}
	\begin{tabular}{ccll}
                $j(J)$ & $\equiv$ & $2J$  & spin-$\frac{1}{2}$-particles, \\
                $u(J,M)$ & $\equiv$ & $J+M$ & spin-up particles
                {\scriptsize$\nearrow$}, \\
                $d(J,M)$ & $\equiv$ & $J-M$ & spin-down particles
                {\scriptsize$\searrow$}.
	\end{tabular}
	\end{center}
This idea relies on the fact that the $2J$ spin-$\frac{1}{2}$-particles
satisfy the same transformation relations under rotation as a
spin-$J$-particle. Thus, if we are interested in the angular momentum of a
spin-$J$-particle, i.e. its transformation properties under rotation,
we might consider $j=2J$ spin-$\frac{1}{2}$-particles as well.
Only in this sense we might speak of a spin-$J$-particles consisting
of spin-$\frac{1}{2}$-particles.
In order to distinguish real spin-$\frac{1}{2}$-particles from the
constituent spin-$\frac{1}{2}$-particles we will call the latter
{\em imaginary spin-$\frac{1}{2}$-particles}.

In the following the question, how many possibilities of decomposing
a particle described by the state $\vert J, M \rangle$ into
imaginary spin-$\frac{1}{2}$-particles exist (see table \ref{pasc1}), will
play a crucial role. We know that there are $u=J+M$ imaginary particles
of the $j$ particles after the decomposition in the state
{\scriptsize$\nearrow$}. Then the number of possibilities is given
by
	\begin{equation}
                s(j,u) \equiv {j \choose u} = \frac{j!}{u!d!}.
	\end{equation}
We say, the state $\vert J, M \rangle$ consists of $s$ {\em imaginary
states}. As we will see later on, the CG-coefficients depend on the
number of the imaginary states belonging to the particles under
consideration. This number is given by the $2J$th row of the binominal
expansion (Pascal's triangle)
(see table \ref{pasc2}) for all (for the spin-$J$-particle) possible
$M$-values.


\subsection{Computation of the CG-coefficients for $J=J_{1}+J_{2}$}

\subsubsection*{The matrix ${\bf \Omega}_{0}$}

We will now investigate the special case of the coupling of two particles
with angular momenta $J_{1}$ and $J_{2}$ to a system with the maximum
angular momentum
		\begin{displaymath}
                        J=J_{1}+J_{2},\ j=j_{1}+j_{2}.
		\end{displaymath}
%
As we
have seen in (\ref{trans}), the coupled state $\vert J,M \rangle$ is
composed of all uncoupled states $\vert (M_{1},M_{2}) \rangle$ for which
$M=M_{1}+M_{2}$ holds. Since there must be no interference between the
uncoupled states -- this would lead to $J<J_{1}+J_{2}$ -- , we expect
real and positive CG-coefficients (according to the usual phase convention).
This enables us to determine the CG-coefficients from square roots of
probabilities. The square of the CG-coefficient
$\langle (M_{1},M_{2}) \vert J,M \rangle$ corresponds to the probability
of finding the individual particles in the state $\vert (M_{1},M_{2})
\rangle$ at the instance of measuring the total angular momentum.
This probability follows by decomposing the coupled and uncoupled states
according to section \ref{abschnitt_zerlegen} into their imaginary
spin-$\frac{1}{2}$-particles and counting the number of imaginary states
generated in this way. We denote the number of imaginary states
of the coupled system by $s_{c}$ and the one of the uncoupled system
by $s_{uc}$, respectively.
   \begin{eqnarray*}
    s_{c}(j,u) & = & {j \choose u},  \\
    s_{uc}(j_{1},u_{1},j_{2},u_{2}) & = & {j_{1} \choose u_{1}}{j_{2}
    \choose u_{2}}.
   \end{eqnarray*}
The quantity $s_{uc}$ is given by the product
of the number of the imaginary states of the individual particles, because
each of the ${j_{1} \choose u_{1}}$ states of particle 1 can generate
a new state with each of the ${j_{2} \choose u_{2}}$ states of particle 2.
The fact that the coupled state and the sum of all uncoupled states with
$M=M_{1}+M_{2}$ describe the same physical system suggests that
both should have the same number of imaginary states, leading to
		\begin{equation}
                 s_c(j,u)= \sum_{u_{1},u_{2}}
                 s_{uc}(j_1,u_1,j_2,u_2),
			\label{kombinatorik}
		\end{equation}
where the sums extends over all $u_{1},u_{2}$ with $u_{1}+u_{2}=u$,
corresponding to $M_{1}+M_{2}=M$. Eq. (\ref{kombinatorik}) is known
as the addition theorem for binomial coefficients  (see e.g. \cite{bronst}).

\begin{quote}
The desired probabilities, i.e. the squares of the CG-coefficients, will
be obtained by the simple assumption that the combined system is found
in each of the ${j \choose u}$ imaginary states with equal probability.
\end{quote}
If we could determine
the imaginary state of the system, we would obtain a state after ${j_{1}
\choose u_{1}}{j_{2} \choose u_{2}}$ of ${j \choose u}$ measurements
in the average,
which according to (\ref{kombinatorik}) originates uniquely from the
decomposition of the uncoupled state $\vert (M_{1},M_{2}) \rangle$.
Hence the desired propability of finding the system in the state $\vert
(M_{1},M_{2}) \rangle$ is given by ${j_{1} \choose u_{1}}{j_{2}
\choose u_{2}}{j \choose u}^{-1}$. In this way we obtain an expression
for the CG-coefficient in the case $J=J_{1}+J_{2}$:
 \begin{eqnarray} \label{cgmax}
 \langle (M_{1},M_{2}) \vert J,M \rangle
 & = & \sqrt{{j_{1} \choose u_{1}}{j_{2} \choose u_{2}}{j \choose u}^{-1}} \\
 & = & \sqrt{\frac{s_{uc}(j_{1},u_{1},j_{2},u_{2})}{s_{c}(j,u)}}. \nonumber
 \end{eqnarray}
Therefore, in this special case the CG-coefficients can be understood in an
elementary way. Nature chooses the simplest way, not distinguishing any of
the $s_{c}$ imaginary states. The $u=u_{1}+u_{2}$
{\scriptsize $\nearrow$}-states
and the $d=d_{1}+d_{2}$ {\scriptsize$\searrow$}-states are coupled with
equal probabilities to all possible imaginary
states. Some of these states ($s_{uc}$),
however, are interpreted as an uncoupled state. Thus the coupled system
consists according to the ratio $\frac{s_{uc}}{s_{c}}$ of this
uncoupled state.

Expressing (\ref{cgmax}) by $J_{i},M_{i}$ and using the definition of the
binomial coefficients, we recover the Racah formula for $J=J_{1}+J_{2}$,
		\begin{eqnarray}  \label{special_racah}
			\lefteqn{\langle (M_{1},M_{2}) \vert J,M \rangle =
			\sqrt{\frac{(2J_{1})!(2J_{2})!}{(2J)!}}} \\
			& & \times\sqrt{\frac{(J+M)!(J-M)!}
                        {(J_{1}+M_{1})!(J_{1}-M_{1})!
                        (J_{2}+M_{2})!(J_{2}-M_{2})!}}, \nonumber
		\end{eqnarray}
as it can be found e.g. in Ref.\cite{mess}.

In order to simplify the explicit calculation of CG-coefficients and the
generalization to arbitrary total angular momenta, we define
a matrix ${\bf \Omega}_{0}$ containing all essential information of the
r.h.s. of (\ref{cgmax}) for all $u_{1},u_{2}$. The crucial quantities
are the values of $s_{uc}$. The values of
$s_{c}$ only take care for the correct
normalization of the CG-coefficients and depend according to
(\ref{kombinatorik}) on $s_{uc}$. We define the matrix ${\bf \Omega}_{0}$
via their components
		\begin{equation}
			{\bf \Omega}_{0} \vert_{u_{1},u_{2}} \equiv
                        s_{uc}(j_{1},u_{1},j_{2},u_{2}) =
                        {j_{1} \choose u_{1}}{j_{2} \choose u_{2}}.
			\label{def_omega0}
		\end{equation}
Here  $u_{1}$ extends from $0$ (first row) to $j_{1}$ and $u_{2}$ from $0$
(first column) to $j_{2}$, respectively. ${\bf \Omega}_{0}$ is a
$2J_{1}+1 \times 2J_{2}+1$-matrix. Sometimes we will use the angular momenta
$J_{1},J_{2}$ as the arguments of ${\bf \Omega}_{0}=
{\bf \Omega}_{0}(J_{1},J_{2})$.


\subsubsection*{Example: Addition of two spin-1-particles to spin 2}

		\label{beispiel_oneone}

One of the essential advantages of (\ref{cgmax}) compared with the Racah
formula (\ref{special_racah}) consists in the possibility of extracting
all results of (\ref{cgmax}) easily by means of the matrix
${\bf \Omega}_{0}$.
As a first example we will calculate all CG-coefficients corresponding to the
coupling of two spin-1-particles to a spin-2-state according to
(\ref{cgmax}).
${\bf \Omega}_{0}$ follows according to (\ref{def_omega0}) by multiplying
two $J=1$-rows of Pascal's triangle, denoted in the following way
		\begin{displaymath}
			{\bf \Omega}_{0}(1,1) =
			\left(
			\begin{array}{c|ccc}
				  & 1 & 2 & 1 \\
			\hline
				1 & 1 & 2 & 1 \\
				2 & 2 & 4 & 2 \\
				1 & 1 & 2 & 1
			\end{array}
			\right)
			=
			\left(
			\begin{array}{ccc}
				1 & 2 & 1 \\
				2 & 4 & 2 \\
				1 & 2 & 1
			\end{array}
                        \right).
		\end{displaymath}
Each of the $3\times3$ components of this matrix reproduces the number
of the imaginary states of the state $\vert (M_{1},M_{2}) \rangle$.
The connection between the position in the matrix
and the corresponding $M$ or $u$-value is shown in table \ref{oneone}.
The maximum state $\vert (1,1) \rangle$ is found down-right.
The diagonals (raising from left to right) correspond to a given
$M_{1}+M_{2}=M$. The $M$-values of the diagonals increase from
$M=-J_{1}-J_{2}$ (up-left) to $M=J_{1}+J_{2}$ (down-right).
The quantity ${j \choose u}$ appearing in (\ref{cgmax}) is extracted
according to (\ref{kombinatorik}) by summing up the diagonal
elements. In our example these sums give $\{1,4,6,4,1\}$, corresponding
to the $J=2$-row of Pascal's triangle, as expected from (\ref{kombinatorik}).
The CG-coefficients are read off from ${\bf \Omega}_{0}$ by dividing the
components ${j_{1} \choose u_{1}}{j_{2} \choose u_{2}}$ by the sum
of the diagonal elements ${j \choose u}$ and extracting the square root.
This procedure yields the following CG-coefficients, as can also be found
in Ref.\cite{cgtables}
  \begin{eqnarray*}
   \vert 2, 2 \rangle & = & \vert (1, 1) \rangle, \\
   \vert 2, 1 \rangle & = & \sqrt{\frac{2}{4}}\vert (1,0) \rangle
   +\sqrt{\frac{2}{4}}\vert (0,1) \rangle, \\
   \vert 2, 0 \rangle & = & \sqrt{\frac{1}{6}}\vert (1,-1) \rangle
   +\sqrt{\frac{4}{6}}\vert (0,0) \rangle +\sqrt{\frac{1}{6}}\vert (-1,1)
   \rangle,\\
   \vert 2, -1 \rangle & = & \sqrt{\frac{2}{4}}\vert (0,-1) \rangle
   +\sqrt{\frac{2}{4}}\vert (-1,0) \rangle, \\
   \vert 2, -2 \rangle & = & \vert (-1, -1) \rangle .
  \end{eqnarray*}
Utilizing the ${\bf \Omega}_{0}$-matrices we are able to determine
all non-vanishing CG-coefficients with $J=J_1+J_2$ immediately. As a further
example we present the coupling of a $J_{1}=\frac{3}{2}$-particle
with a $J_{2}=1$-particle to the $J=\frac{5}{2}$-state:
		\begin{displaymath}
			{\bf \Omega}_{0}(\frac{3}{2},1) = \left(
			\begin{array}{c|ccc}
				  & 1 & 2 & 1 \\
			\hline
				1 & 1 & 2 & 1 \\
				3 & 3 & 6 & 3 \\
				3 & 3 & 6 & 3 \\
				1 & 1 & 2 & 1
			\end{array}
			\right)
			=
			\left(
			\begin{array}{ccc}
				1 & 2 & 1 \\
				3 & 6 & 3 \\
				3 & 6 & 3 \\
				1 & 2 & 1
			\end{array}
                        \right).
		\end{displaymath}
%

\section{General addition of two angular momenta}

Next we will consider the general case in which two particles with
$J_{1}$ and $J_{2}$ may couple to a total angular momentum
$J<J_{1}+J_{2}$. This problem can also be treated by means of imaginary
spin-$\frac{1}{2}$-particles. Assuming that two of these imaginary
paricles couple to spin 0, they do not contribute to the total
angular momentum $J$ anymore. If there are $n$ spin-0-particles
in a system of imaginary spin-$\frac{1}{2}$-particles, the
total angular momentum is reduced to $J=J_{1}+J_{2}-n$. In this way
the general addition of two angular momenta can be described by two
elementary processes of imaginary spin-$\frac{1}{2}$-particles, namely
the coupling of some imaginary particles to spin 0 and the coupling
of the remaining particles to their maximum angular momentum according
to section II. First we will consider the coupling of two
spin-$\frac{1}{2}$-particles to spin 0, before we will combine both
processes.

\newpage

\subsubsection*{Coupling of two imaginary spin-$\frac{1}{2}$-particles to
spin 0}

{}From
                \begin{displaymath}
			{\bf \Omega}_{0}(\frac{1}{2},\frac{1}{2}) =
			\left(
				\begin{array}{c|cc}
					1 & 1 & 1 \\
					\hline
					1 & 1 & 1 \\
					1 & 1 & 1
				\end{array}
			\right)
			=
			\left(
				\begin{array}{cc}
					1 & 1 \\
					1 & 1 \\
				\end{array}
			\right)
		\end{displaymath}
we obtain
        \begin{displaymath}
		\vert 1, 0 \rangle = \sqrt{\frac{1}{2}} \vert
		(\frac{1}{2},-\frac{1}{2}) \rangle + \sqrt{\frac{1}{2}} \vert
		(-\frac{1}{2},\frac{1}{2}) \rangle.
	\end{displaymath}
{}From the requirement $\langle (0,0) \vert (1,0) \rangle=0$
we find
	\begin{displaymath}
		\vert 0, 0 \rangle = \sqrt{\frac{1}{2}} \vert
		(\frac{1}{2},-\frac{1}{2}) \rangle - \sqrt{\frac{1}{2}} \vert
		(-\frac{1}{2},\frac{1}{2}) \rangle.
        \end{displaymath}
We indicate the state $\vert 0, 0 \rangle$ as
        \begin{equation}
		\nearrow\searrow - \searrow\nearrow.
		\label{kill}
	\end{equation}
%

\subsection{The matrix ${\bf \Omega}_{n}$ of the coupling to
$J=J_{1}+J_{2}-n$}
	\label{abschnitt_omegan}

As we will see, the general case $J=J_{1}+J_{2}-n$ can also be constructed
from a matrix, called ${\bf \Omega}_{n}$
from which we can read off the CG-coefficient,
similar to ${\bf \Omega}_{0}$, directly. However, we cannot
demand positiveness of the CG-coefficients any longer. Negative
CG-coefficients will be identified from ${\bf \Omega}_{n}$ simply by
a minus sign in front of the corresponding component.

In order to exemplify this prescription we will show
${\bf \Omega}_{1}(\frac{3}{2},1)$ here. The details of its calculation are
given below. It describes the coupling of
a spin-$\frac{3}{2}$-particle with a spin-$1$-particle to spin $\frac{3}{2}$:

		\begin{displaymath}
			{\bf \Omega}_{1}(\frac{3}{2},1) = \left(
			\begin{array}{ccc}
				0 & -3 & -6 \\
				2 & -1 & -8 \\
				8 & 1 & -2 \\
				6 & 3 & 0
			\end{array}
                        \right).
		\end{displaymath}
{}From this we can read off the CG-coefficients (from down-left to up-right),
e.g. from the 4th diagonal corresponding to $M=\frac{1}{2}$:
	\begin{displaymath}
         \vert \frac{3}{2}, \frac{1}{2} \rangle = \sqrt{\frac{6}{15}} \vert
		(\frac{3}{2},-1) \rangle + \sqrt{\frac{1}{15}} \vert
		(\frac{1}{2},0) \rangle - \sqrt{\frac{8}{15}} \vert
                (-\frac{1}{2},1) \rangle .
	\end{displaymath}
The minus signs in ${\bf \Omega}_{1}$ just indicate that the corresponding
CG-coefficient is negative. Calculating the normalization, these signs
must not be considered in the sum of the diagonal elements, here 15.

We will regard two ${\bf \Omega}_{n}$ matrices as equivalent, if they differ
only by a factor $\alpha$, denoted by $\alpha{\bf \Omega}_{n}
\hat={\bf \Omega}_{n}$. Also two matrices are equivalent, if their
diagonals, raising from left to right, deviate only by a factor.
In both cases we find the same CG-coefficients.

\begin{quote}
The matrix ${\bf \Omega}_{n}$ follows from the requirements that
\begin{enumerate}
\item among the imaginary spin-$\frac{1}{2}$-particles $n$ Spin-0-particles
arise
                \begin{center} {\em and} \end{center}
\item the remaining imaginary spin-$\frac{1}{2}$-particles couple to their
maximum spin.
\end{enumerate}
Spin-0-particles arise according to (\ref{kill}), if the imaginary
spin-$\frac{1}{2}$-particles generate
\begin{center}
{\scriptsize$\nearrow\searrow$}- \hspace{1cm}{\em or}
\hspace{1cm}{\scriptsize$\searrow\nearrow$}-states.
\end{center}
\end{quote}

The components of ${\bf \Omega}_{n}$ play the role of non-normalized
probabilities, analogously to the interpretation of ${\bf \Omega}_{0}$.
They are composed from the probabilities of the processes 1. and 2.
Defining
\begin{itemize}
\item  ${\bf\Lambda}$ as the matrix describing the probabilities
for the state {\scriptsize $ \nearrow\searrow$},
\item  ${\bf V}$ as the matrix describing the probabilities for the state
{\scriptsize$\searrow\nearrow$}, and
\item  ${\bf \tilde\Omega}_{n}$ as the matrix describing the probabilities
for the coupling of the remaining imaginary particles to their maximum
spin,
\end{itemize}
the above requirements translates directly to the matrix equation
\begin{equation}
       {\bf \Omega}_{n} = ({\bf\Lambda-V})^{[n]} {\bf \tilde\Omega}_{n},
		\label{omegan1}
\end{equation}
where the products ({\em and}) and the subtractions ({\em or}) between
the matrices have to be performed for each single component. When we
talk of probabilities in the following, we may not refer to normalized
probabilities, as the product by components in (\ref{omegan1})
allows for arbitrary factors (normalization) in ${\bf\Lambda-V}$ and ${\bf
\tilde\Omega}_{n}$.


\subsection{The matrix $({\bf\Lambda-V})^{[n]}$ of $n$ spin-0-particles}
	\label{sec_lambdav}

For calculating $({\bf\Lambda-V})^{[n]}$ we need the probabilities
of creating {\scriptsize $\nearrow\searrow$}- or
{\scriptsize $\searrow\nearrow$}-states from imaginary
spin-$\frac{1}{2}$-particles. The probability of creating
{\scriptsize $\nearrow\searrow$} is proportional to the probability of
finding {\scriptsize$\nearrow$} at particle 1 and {\scriptsize$\searrow$}
at particle 2 at the same time. This in turn is proportional to the
product of the
number of the imaginary particles in the corresponding states. An analogous
statement holds for {\scriptsize$\searrow\nearrow$}.  Therefore we define
\begin{eqnarray}
 {\bf\Lambda} \vert_{u_{1},u_{2}} & \equiv & u_{1}d_{2} = u_{1}(j_{2}-u_{2})
                \label{def_lambda_v},  \\
 {\bf V} \vert_{u_{1},u_{2}} & \equiv & d_{1}u_{2} = (j_{1}-u_{1})u_{2},
 \nonumber
\end{eqnarray}
leading to
\[ ({\bf\Lambda-V})|_{u_{1},u_{2}}=u_{1}d_{2}-d_{1}u_{2}. \]
The expression $u_{1}d_{2}-d_{1}u_{2}$ can be interpreted (if it is positive)
as the surplus of {\scriptsize$\nearrow\searrow$}-states of an uncoupled
state. If it is negative it yields the (negative) excess of
{\scriptsize$\searrow\nearrow$}-states. This pays attention to the fact
that according to (\ref{trans}) interference can occur only between
uncoupled states but not within an uncoupled state itself.

Performing the exponentiation $[n]$ one has to take into account that
the probabilities ${\bf\Lambda-V}$ depend on the number of the
spin-0-particles already generated. If e.g. already one
{\scriptsize $\nearrow\searrow$}-state has been built, there are only
$u_{1}-1$ and $d_{2}-1$ particles available. Therefore the
exponentiation means
	\begin{equation}
               ({\bf\Lambda-V})^{[n]}|_{u_{1},u_{2}} \equiv
               \sum_{k=0}^{n}(-)^{k}{n \choose
		k}(u_{1}d_{2})^{[n-k]}(d_{1}u_{2})^{[k]}
                \label{lamiv},
	\end{equation}
where
	\begin{eqnarray}
                (u_{1}d_{2})^{[n-k]} & \equiv &
                \prod_{i=0}^{n-k-1}(u_{1}-i)(d_{2}-i), \nonumber  \\
                (d_{1}u_{2})^{[k]} & \equiv &
                \prod_{i=0}^{k-1}(d_{1}-i)(u_{2}-i). \label{exponent}
	 \end{eqnarray}
%

\subsubsection*{Example: $({\bf\Lambda-V})^{[n]}$ for $J_{1}=\frac{3}{2}$
and $J_{2}=1$}

As an example we will calculate ${\bf\Lambda-V}$ and $({\bf\Lambda-V})^{[2]}$
for the coupling of $J_{1}=\frac{3}{2}$ and $J_{2}=1$, already mentioned
above. The ${\bf \Lambda}$ and ${\bf V}$ matrices in this case read
	\begin{displaymath}
		{\bf \Lambda} = \left(
		\begin{array}{c|ccc}
		  & 2 & 1 & 0 \\
		 \hline
		0 & 0 & 0 & 0 \\
		1 & 2 & 1 & 0 \\
		2 & 4 & 2 & 0 \\
		3 & 6 & 3 & 0
		\end{array}
		\right)
		\text{, }
		{\bf V} = \left(
		\begin{array}{c|ccc}
		  & 0 & 1 & 2 \\
		 \hline
		3 & 0 & 3 & 6 \\
		2 & 0 & 2 & 4 \\
		1 & 0 & 1 & 2 \\
		0 & 0 & 0 & 0
		\end{array}
                \right),
	\end{displaymath}
leading to
        \begin{displaymath}
		{\bf\Lambda-V} = \left(
		\begin{array}{ccc}
		0 & -3 & -6 \\
		2 & -1 & -4 \\
		4 & 1 & -2 \\
		6 & 3 & 0
                \end{array} \right).
	\end{displaymath}
The matrix $({\bf\Lambda-V})^{[2]}$ is given according to (\ref{lamiv})
and (\ref{exponent}) by

        \[ ({\bf\Lambda-V})^{[2]}={\bf\Lambda \Lambda'}
        -2{\bf \Lambda V}+{\bf V V'}, \]
where
        \[ {\bf \Lambda'} \vert_{u_{1},u_{2}} = (u_{1}-1)(d_{2}-1) \]
and
        \[ {\bf V'} \vert_{u_{1},u_{2}}=(d_{1}-1)(u_{2}-1) \]
read
  	\begin{displaymath}
		{\bf \Lambda'} = \left(
		\begin{array}{c|ccc}
			  & 1 & 0 & 0 \\
			 \hline
			0 & 0 & 0 & 0 \\
			0 & 0 & 0 & 0 \\
			1 & 1 & 0 & 0 \\
			2 & 2 & 0 & 0
		\end{array}
		\right)
		\text{, }
		{\bf V'} = \left(
		\begin{array}{c|ccc}
			  & 0 & 0 & 1 \\
			 \hline
			2 & 0 & 0 & 2 \\
			1 & 0 & 0 & 1 \\
			0 & 0 & 0 & 0 \\
			0 & 0 & 0 & 0
		\end{array}
                \right).
	\end{displaymath}
{}From this we get
	\begin{eqnarray*}
		({\bf\Lambda-V})^{[2]} & = & \left(
		\begin{array}{ccc}
			0 & 0 & 0 \\
			0 & 0 & 0 \\
			4 & 0 & 0 \\
			12 & 0 & 0
		\end{array} \right)
		-2 \left(
		\begin{array}{ccc}
			0 & 0 & 0 \\
			0 & 2 & 0 \\
			0 & 2 & 0 \\
			0 & 0 & 0
		\end{array} \right)
		+ \left(
		\begin{array}{ccc}
			0 & 0 & 12 \\
			0 & 0 & 4 \\
			0 & 0 & 0 \\
			0 & 0 & 0
		\end{array} \right) \\
		& = & \left(
		\begin{array}{ccc}
			0 & 0 & 12 \\
			0 & -4 & 4 \\
			4 & -4 & 0 \\
			12 & 0 & 0
		\end{array} \right)
		\hat= \left(
		\begin{array}{ccc}
			0 & 0 & 3 \\
			0 & -1 & 1 \\
			1 & -1 & 0 \\
			3 & 0 & 0
		\end{array} \right).
	\end{eqnarray*}
All higher exponents ($n=3,4,...$) of ${\bf\Lambda-V}$ vanish in
accordance with the rule $\vert J_{1}-J_{2} \vert \le J \le J_{1}+J_{2}$.
In order to calculate the CG-coefficients of the above example, we need
the matrices ${\bf \tilde\Omega}_{1}$ and ${\bf \tilde\Omega}_{2}$
of the remaining system in addition. Their computation is the topic
of the next section.


\subsection{The matrix ${\bf \tilde\Omega}_{n}$ of the remaining system}

	\label{sec_tildeomega}

The matrix ${\bf \tilde\Omega}_{n}$ follows directly from ${\bf \Omega}_{0}$
by removing the imaginary spin-$\frac{1}{2}$-particles, contained in
$({\bf\Lambda-V})^{[n]}$. Here removing a {\scriptsize$\nearrow$}
from the particle $i$ means that the total number of imaginary
spin-$\frac{1}{2}$-particles and the number of {\scriptsize$\nearrow$}
is reduced by one; i.e., we replace
	\begin{eqnarray*}
                j_{i} & \mapsto & j_{i}-1, \\
		u_{i} & \mapsto & u_{i}-1.
	\end{eqnarray*}
Analogously a {\scriptsize$\searrow$} is removed by
	\begin{eqnarray*}
                j_{i} & \mapsto & j_{i}-1, \\
		d_{i} & \mapsto & d_{i}-1.
	\end{eqnarray*}
For constructing ${\bf \tilde\Omega}_{1}$ a spin-0-particle has to be removed
from ${\bf \Omega}_{0}$; i.e., according to (\ref{kill})
a {\scriptsize$\nearrow$} will be removed from particle 1 and a
{\scriptsize$\searrow$} from particle 2 at the same time {\em or} a
{\scriptsize$\searrow$} from particle 1 and a {\scriptsize$\nearrow$}
from particle 2. Together with

	\[
        {\bf \Omega}_{0}|_{u_{1},u_{2}} = {j_{1} \choose u_{1}}{j_{2}
        \choose u_{2}}
	\]
we get
	\[
	{\bf \tilde\Omega}_{1}|_{u_{1},u_{2}} \equiv {j_{1}-1 \choose
        u_{1}-1}{j_{2}-1 \choose u_{2}} - {j_{1}-1 \choose u_{1}}{j_{2}-1
        \choose u_{2}-1},
        \]
where the removing of a {\scriptsize$\searrow$} has been performed via
	\[
	{j_{i} \choose u_{i}}={j_{i} \choose d_{i}} \mapsto {j_{i}-1 \choose
        d_{i}-1} = {j_{i}-1 \choose u_{i}}.
	\]
The binomial coefficients vanish for all unphysical values, e.g.
if $j_{i}-1<u_{i}$ or $u_{i}-1<0$.
This corresponds to the fact that there is no state with $u_{i}<0$ or
$d_{i}<0$. The matrix  ${\bf \tilde\Omega}_{2}$ is obtained from
${\bf \tilde\Omega}_{1}$ by removing an additional spin-0-particle:
	\begin{eqnarray*}
		{\bf \tilde\Omega}_{2}|_{u_{1},u_{2}} & = & {j_{1}-2 \choose
		u_{1}-2}{j_{2}-2 \choose u_{2}} \\
		& & - 2{j_{1}-2 \choose u_{1}-1}{j_{2}-2 \choose u_{2}-1} \\
                & & +  {j_{1}-2 \choose u_{1}}{j_{2}-2 \choose u_{2}-2}.
	\end{eqnarray*}
In general ${\bf \tilde\Omega}_{n}$ is given by
	\begin{equation}\label{def_tildeomega}
                {\bf \tilde\Omega}_{n}|_{u_{1},u_{2}} \equiv \sum_{k=0}^{n}
                (-)^{k} {n \choose k}
                {j_{1}-n \choose u_{1}-(n-k)}{j_{2}-n \choose u_{2}-k}.
	\end{equation}
The matrices ${\bf \tilde\Omega}_{n}$ can be calculated easily, as they
consist essentially out of ${\bf \Omega}_{0}(J_{1}-\frac{n}{2},
J_{2}-\frac{n}{2})$. This reflects the fact that a system with
$J=J_{1}+J_{2}-n$ can be imagined as originated from coupling of
spin $J_{1}-\frac{n}{2}$ and spin $J_{2}-\frac{n}{2}$ to their
maximum spin. This will be illuminated in the following example.


\subsubsection*{Example: ${\bf \tilde\Omega}_{n}$ for $J_{1}=\frac{3}{2}$
and $J_{2}=1$}

We compute ${\bf \tilde\Omega}_{1}$ and ${\bf \tilde\Omega}_{2}$ in the
case of the coupling of $J_{1}=\frac{3}{2}$ and $J_{2}=1$ from
(\ref{def_tildeomega}). First, however, we will repeat ${\bf \Omega}_{0}$:
	\begin{displaymath}
		{\bf \Omega}_{0}(\frac{3}{2},1) = \left(
		\begin{array}{ccc}
		1 & 2 & 1 \\
		3 & 6 & 3 \\
		3 & 6 & 3 \\
		1 & 2 & 1
		\end{array}
                \right).
	\end{displaymath}
For physical values of $u_{1}$ and $u_{2}$ the expression ${j_{1}-n
\choose u_{1}-(n-k)}{j_{2}-n \choose u_{2}-k}$ in (\ref{def_tildeomega})
leads to the smaller matrix ${\bf \Omega}_{0}(J_{1}-\frac{n}{2},
J_{2}-\frac{n}{2})$, in our example
	\begin{displaymath}
		{\bf \Omega}_{0}(1,\frac{1}{2}) = \left(
		\begin{array}{cc}
		1 & 1 \\
		2 & 2 \\
		1 & 1 \\
		\end{array}
                \right)
	\end{displaymath}
if $n=1$ and
	\begin{displaymath}
		{\bf \Omega}_{0}(\frac{1}{2},0) = \left(
		\begin{array}{c}
		1 \\
		1 \\
		\end{array}
                \right)
	\end{displaymath}
if $n=2$. For all other values of $u_{1}$ and $u_{2}$ the expression
is identical to zero. The value of $k$ fixes the position of these
matrices: $k$=0 corresponds to down-left, $k$=$n$ to up-right. The matrices
${\bf \tilde\Omega}_{1}$ and ${\bf \tilde\Omega}_{2}$ thus are found easily:
	\begin{displaymath}
		{\bf \tilde\Omega}_{1} =
		\left(
			\begin{array}{ccc}
			0 & 0 & 0 \\
			1 & 1 & 0 \\
			2 & 2 & 0 \\
			1 & 1 & 0
			\end{array}
		\right)
		-
		\left(
			\begin{array}{ccc}
			0 & 1 & 1 \\
			0 & 2 & 2 \\
			0 & 1 & 1 \\
			0 & 0 & 0
			\end{array}
		\right)
		=
		\left(
			\begin{array}{ccc}
			0 & -1 & -1 \\
			1 & -1 & -2 \\
			2 & 1 & -1 \\
			1 & 1 & 0
			\end{array}
		\right),
	\end{displaymath}

	\begin{eqnarray*}
		{\bf \tilde\Omega}_{2} & = &
		\left(
			\begin{array}{ccc}
			0 & 0 & 0 \\
			0 & 0 & 0 \\
			1 & 0 & 0 \\
			1 & 0 & 0
			\end{array}
		\right)
		-2
		\left(
			\begin{array}{ccc}
			0 & 0 & 0 \\
			0 & 1 & 0 \\
			0 & 1 & 0 \\
			0 & 0 & 0
			\end{array}
		\right)
		+
		\left(
			\begin{array}{ccc}
			0 & 0 & 1 \\
			0 & 0 & 1 \\
			0 & 0 & 0 \\
			0 & 0 & 0
			\end{array}
		\right)
		\\
		& = &
		\left(
			\begin{array}{ccc}
			0 & 0 & 1 \\
			0 & -2 & 1 \\
			1 & -2 & 0 \\
			1 & 0 & 0
			\end{array}
		\right).
	\end{eqnarray*}
%

\subsection{Results of the example $J_{1}=\frac{3}{2}$ and $J_{2}=1$}

	\label{sec_ergebnisse}

We are now able to give all CG-coefficients in form of the
${\bf \Omega}$-matrices ${\bf \Omega}_{0}$ to ${\bf \Omega}_{2}$,
producing the well-known results \cite{cgtables}.

\begin{itemize}
\item Coupling of $J_{1}=\frac{3}{2}$ and $J_{2}=1$ to $J=\frac{5}{2}$:
\end{itemize}
		\begin{displaymath}
			{\bf \Omega}_{0}(\frac{3}{2},1) = \left(
			\begin{array}{ccc}
			1 & 2 & 1 \\
			3 & 6 & 3 \\
			3 & 6 & 3 \\
			1 & 2 & 1
			\end{array}
			\right).
		\end{displaymath}
\begin{itemize}
\item Coupling of $J_{1}=\frac{3}{2}$ and $J_{2}=1$ to $J=\frac{3}{2}$:
\end{itemize}
{}From (\ref{omegan1}) together with the results of the sections
\ref{sec_lambdav} and \ref{sec_tildeomega} we end up with
	\begin{eqnarray*}
          {\bf \Omega}_{1} & = & ({\bf\Lambda-V}) {\bf \tilde\Omega}_{1} \\
		 	& = &
		 	\left(
				\begin{array}{ccc}
				0 & -3 & -6 \\
				2 & -1 & -4 \\
				4 & 1 & -2 \\
				6 & 3 & 0
				\end{array} \right)
			\left(
				\begin{array}{ccc}
				0 & -1 & -1 \\
				1 & -1 & -2 \\
				2 & 1 & -1 \\
				1 & 1 & 0
				\end{array}
			\right)
			=
			\left(
				\begin{array}{ccc}
				0 & -3 & -6 \\
				2 & -1 & -8 \\
				8 & 1 & -2 \\
				6 & 3 & 0
				\end{array}
                        \right).
	\end{eqnarray*}
It should be noticed once more that a minus sign in front of a component
only indicates that the corresponding CG-coefficient is negative. Thus a
minus sign is transferred to the product matrix, although if both the
components which are multiplied are negative.
(As we will see in the next section in (\ref{tildeomega2}), the components
which are multiplied have the same sign.)
The matrix
${\bf \Omega}_{1}$ has been discussed already in section
\ref{abschnitt_omegan}, where the extraction of the CG-coefficients from it
has been demonstrated.

\begin{itemize}
\item Coupling of $J_{1}=\frac{3}{2}$ and $J_{2}=1$ to $J=\frac{1}{2}$:
\end{itemize}
\begin{eqnarray*}
   {\bf \Omega}_{2} & = & ({\bf\Lambda-V})^{[2]} {\bf \tilde\Omega}_{2} \\
		 	& = &
		 	\left(
				\begin{array}{ccc}
				0 & 0 & 3 \\
				0 & -1 & 1 \\
				1 & -1 & 0 \\
				3 & 0 & 0
				\end{array} \right)
			\left(
				\begin{array}{ccc}
				0 & 0 & 1 \\
				0 & -2 & 1 \\
				1 & -2 & 0 \\
				1 & 0 & 0
				\end{array}
			\right)
			=
			\left(
				\begin{array}{ccc}
				0 & 0 & 3 \\
				0 & -2 & 1 \\
				1 & -2 & 0 \\
				3 & 0 & 0
				\end{array}
                        \right).
	\end{eqnarray*}
%

\section{Alternative formulation}

So far we are able to compute all CG-coefficients from (\ref{omegan1})

	\begin{displaymath}
    {\bf \Omega}_{n} = ({\bf\Lambda-V})^{[n]} {\bf \tilde\Omega}_{n}.
	\end{displaymath}
Comparing the definition (\ref{lamiv}) of $({\bf\Lambda-V})^{[n]}$
with the definition (\ref{def_tildeomega}) of ${\bf \tilde\Omega}_{n}$,
the following relation between both the matrices can be recognized:
	\begin{equation}
         {\bf \tilde\Omega}_{n} \hat= ({\bf\Lambda-V})^{[n]}{\bf \Omega}_{0}.
		\label{tildeomega2}
	\end{equation}
This relation
offers the possibility of finding an alternative formulation that requires
only the knowledge of one of the both matrices and ${\bf \Omega}_{0}$.
It can be shown directly by multiplying $({\bf\Lambda-V})^{[n]}$ with
${\bf \Omega}_{0}$. For this purpose the terms (\ref{exponent}) in
(\ref{lamiv}) are written as
	 \begin{eqnarray*}
                (u_{1}d_{2})^{[n-k]} & = &
                \frac{u_{1}!d_{2}!}{(u_{1}-(n-k))!(d_{2}-(n-k))!},  \\
                (d_{1}u_{2})^{[k]}& = &
                \frac{d_{1}!u_{2}!}{(d_{1}-k)!(u_{2}-k)!}.
	 \end{eqnarray*}
Multiplying these terms with the components of ${\bf \Omega}_{0}$
	\[
        {j_{1} \choose u_{1}}{j_{2} \choose u_{2}}=
        \frac{j_{1}!j_{2}!}{u_{1}!d_{1}!u_{2}!d_{2}!}
	\]
yields
\begin{eqnarray*}
\lefteqn{(u_{1}d_{2})^{[n-k]}(d_{1}u_{2})^{[k]}{j_{1}
\choose u_{1}}{j_{2} \choose u_{2}}}\\
& = & \frac{j_{1}!}{(u_{1}-(n-k))!(d_{1}-k)!}
\frac{j_{2}!}{(u_{2}-k)!(d_{2}-(n-k))!}.
\end{eqnarray*}
This expression is proportional to
\begin{eqnarray*}
\lefteqn{{j_{1}-n \choose u_{1}-(n-k)}{j_{2}-n \choose u_{2}-k}}\\
& = & \frac{(j_{1}-n)!}{(u_{1}-(n-k))!(d_{1}-k)!}
\frac{(j_{2}-n)!}{(u_{2}-k)!(d_{2}-(n-k))!},
\end{eqnarray*}
showing up in (\ref{def_tildeomega}). Thus (\ref{tildeomega2}) is proven,
from which the following equations, equivalent to (\ref{omegan1}),
result:
	\begin{eqnarray}
          {\bf \Omega}_{n} & \hat= & {\bf \tilde\Omega}_{n}^{2}
          {\bf \Omega}_{0}^{-1},
          \label{omegan2}  \\
          {\bf \Omega}_{n} & \hat= & \left( ({\bf\Lambda-V})^{[n]} \right)^{2}
          {\bf \Omega}_{0}.
          \label{omegan3}
	\end{eqnarray}
Here again we have to take care of the signs when squaring a matrix. The
matrix ${\bf \Omega}_{0}^{-1}$ contains the inverse components of
${\bf \Omega}_{0}$ (or a multiple of them).

As we will show in the appendix, (\ref{omegan2}) and therefore also
(\ref{omegan1}) are equivalent to the Racah formula.


\subsubsection*{Example}

We check (\ref{omegan2}) using the example ${\bf \Omega}_{2}(\frac{3}{2},1)$.
In section \ref{sec_tildeomega} we found
	\begin{displaymath}
		{\bf \tilde\Omega}_{2} =
		\left(
			\begin{array}{ccc}
			0 & 0 & 1 \\
			0 & -2 & 1 \\
			1 & -2 & 0 \\
			1 & 0 & 0
			\end{array}
		\right)
	\Rightarrow
		{\bf \tilde\Omega}_{2}^{2} =
		\left(
			\begin{array}{ccc}
			0 & 0 & 1 \\
			0 & -4 & 1 \\
			1 & -4 & 0 \\
			1 & 0 & 0
			\end{array}
		\right).
	\end{displaymath}
Furthermore,
	\begin{displaymath}
		{\bf\Omega}_{0} =
		\left(
			\begin{array}{ccc}
			1 & 2 & 1 \\
			3 & 6 & 3 \\
			3 & 6 & 3 \\
			1 & 2 & 1
			\end{array}
		\right)
		\Rightarrow
			{\bf\Omega}_{0}^{-1} \hat=
		\left(
			\begin{array}{ccc}
			6 & 3 & 6 \\
			2 & 1 & 2 \\
			2 & 1 & 2 \\
			6 & 3 & 6
			\end{array}
		\right).
	\end{displaymath}
The final result reads
	\begin{eqnarray*}
                 {\bf \Omega}_{2} & \hat= & {\bf \tilde\Omega}_{2}^{2}
                 {\bf \Omega}_{0}^{-1} \\
	 	& = &
	 	\left(
			\begin{array}{ccc}
				0 & 0 & 1 \\
				0 & -4 & 1 \\
				1 & -4 & 0 \\
				1 & 0 & 0
			\end{array} \right)
		\left(
			\begin{array}{ccc}
				6 & 3 & 6 \\
				2 & 1 & 2 \\
				2 & 1 & 2 \\
				6 & 3 & 6
			\end{array}
		\right)
		=
		\left(
			\begin{array}{ccc}
			0 & 0 & 6 \\
			0 & -4 & 2 \\
			2 & -4 & 0 \\
			6 & 0 & 0
			\end{array}
		\right),
	\end{eqnarray*}
which is equivalent to the result of section \ref{sec_ergebnisse} up
to a factor. In conclusion, (\ref{omegan2}) provides a very simple
method for a practicable computation of CG-coefficients. For obtaining
${\bf \tilde\Omega}_{n}^{2}$ and ${\bf \Omega}_{0}^{-1}$ only the
matrices ${\bf \Omega}_{0}(J_{1},J_{2})$ and
${\bf \Omega}_{0}(J_{1}-\frac{n}{2},J_{2}-\frac{n}{2})$ are needed.
Actually, the calculation can be simplified further by considering the
reflection symmetry of ${\bf \Omega}_0$ and ${\bf \Omega}_n$ at their
center, if $n$ is even. If $n$ is odd, the signs of the reflected
components have to be changed. This property of ${\bf \Omega}_0$
and ${\bf \Omega}_n$ corresponds to the relation $\langle (M_1,M_2)\vert
J,M\rangle =(-)^{J_1+J_2-J}\langle (-M_1,-M_2)\vert J,-M\rangle $
\cite{mess}.
Because of the component-like multiplication and subtraction of the matrices
yielding ${\bf \Omega}_n$, only half
of the components of ${\bf \Omega}_0(J_1,J_2)$
and ${\bf \Omega}_0(J_1-\frac{n}{2},J_2-\frac{n}{2})$ has to be computed.


\section{Conclusions}

In this paper we have demonstrated how the coupling of two angular momenta,
$J_1$ and $J_2$,
can be understood easily by decomposing each of the angular momenta into
imaginary spin-$\frac{1}{2}$-particles. Counting the possibilities
of decomposing the angular momenta $J_i$ into these imaginary particles
by simple combinatorial manipulations
(binomial coefficients), leads to the CG-coefficients in the case of
coupling to maximum angular momentum $J=J_1+J_2$.
Here the basic assumption of equally distributed probabilities for each
of the decompositions has been adopted.
For extracting the CG-coefficients,
it is convenient to introduce the matrix ${\bf \Omega}_0$, defined
in (\ref{def_omega0}). The CG-coefficients can be read off directly
from the components of this matrix divided by the sum of the diagonal
elements and extracting the square root.

The general case, $J=J_1+J_2-n$, can be considered as the coupling of
$2n$ imaginary spin-$\frac{1}{2}$-particles to spin 0 and the coupling
of the remaining imaginary particles to maximum spin. In this way the
matrix ${\bf \Omega}_n$ , defined in (\ref{omegan1}), results, from which
the CG-coefficients can be read off in the same way as from
${\bf \Omega}_0$, apart from the fact that minus signs in front of the
components simply indicate negative CG-coefficients. Utilizing a relation
between the matrix describing the coupling of the imaginary particles
to spin 0 and the one describing the coupling of the remaining system to
maximum spin, the matrix ${\bf \Omega}_n$ can be found easily by
(\ref{omegan2}). This method, shown to be equivalent to the Racah
formula, requires much less computational effort than the latter one.

The new, probabilistic interpretation of the addition of angular momenta,
presented
here, provides an elegant and practicable way of computing all
CG-coefficients for the coupling of two arbitrary angular momenta.


\appendix
\section*{Equivalence to the Racah formula}

Here (\ref{omegan2}),
${\bf\Omega}_{n}\hat={\bf\tilde\Omega}_{n}^{2} {\bf\Omega}_{0}^{-1}$,
shall be derived from the Racah formula (see e.g. \cite{mess})
\begin{eqnarray} \label{racahf}
\langle (M_{1}, M_{2})\vert J,M \rangle & = &
\sum_{k}(-)^{k}\sqrt{ (2J+1) }
\sqrt{\frac{(J_{1}+J_{2}-J)!(J_{1}-J_{2}+J)!(J_{2}-J_{1}+J)!}
{(J_{1}+J_{2}+J+1)!(J_{1}-M_{1}-k)^{2}! (J_{2}+M_{2}-k)!^{2} k!^{2}}}
\nonumber\\
& & \times \sqrt{\frac{(J_{1}+M_{1})! (J_{2}+M_{2})!(J+M)!(J_{1}-M_{1})!
(J_{2}-M_{2})!(J-M)!}{(J-J_{2}+M_{1}+k)!^{2}(J-J_{1}-M_{2}+k)!^{2}
(J_{1}+J_{2}-J-k)!^{2}}}.
\end{eqnarray}
For this purpose we express the angular momenta showing up in the Racah
formula by the number of imaginary particles, substituting
\begin{eqnarray*}
        J_{i} & = & j_{i}/2, \\
        J     & = & J_{1}+J_{2}-n = j_{1}/2+j_{2}/2-n, \\
        M_{i} & = & u_{i}-j_{i}/2, \\
        M     & = & M_{1}+M_{2} = u_{1}+u_{2}-j_{1}/2-j_{2}/2.
\end{eqnarray*}
Then we find
\begin{eqnarray} \label{myracah}
        \langle (M_{1}, M_{2})\vert J,M \rangle & = &
        \sum_{k}(-)^{k}{n
        \choose k} \sqrt{\frac{(j+1)u!d!}{(j+1+n)! n!}}  \\
        & & \times \sqrt{\frac{(j_{1}-n)! (j_{2}-n)! u_{1}! d_{1}! u_{2}!
        d_{2}!} {(d_{1}-k)!^{2}
        (u_{1}-n+k)!^{2} (d_{2}-n+k)!^{2}(u_{2}-k)!^{2}} }, \nonumber
\end{eqnarray}
where
\begin{eqnarray*}
u & = & u_1+u_2-n, \\
d & = & d_1+d_2-n, \\
j & = & j_1+j_2-2n.
\end{eqnarray*}
The components ${\bf\Omega}_{n}\vert _{u_1,u_2}$
of the matrix ${\bf\Omega}_{n}$ follow
by squaring the CG-coefficients, taking care of their sign (indicated
by the exponent $|2|$). The first square root in (\ref{myracah})
can be omitted, because it is constant within the diagonals of the matrix.
Thus we get
\begin{equation} \label{myracah2}
        {\bf \Omega}_{n}\vert _{u_1,u_2}
        = \left( \sum_{k}(-)^{k} {n \choose k}
        \sqrt{\frac{(j_{1}-n)! (j_{2}-n)! u_{1}! d_{1}! u_{2}! d_{2}!}
        {(d_{1}-k)!^{2}
        (u_{1}-n+k)!^{2} (d_{2}-n+k)!^{2}(u_{2}-k)!^{2}} } \right)^{|2|}.
\end{equation}
Multiplying (\ref{myracah2}) with the components
\begin{displaymath}
{\bf \Omega}_{0}\vert _{u_1,u_2}={j_{1} \choose u_{1}}{j_{2} \choose
u_{2}}=\frac{j_{1}!}{u_{1}!d_{1}!} \frac{j_{2}!}{u_{2}!d_{2}!}
\end{displaymath}
of the matrix ${\bf\Omega}_{0}$ leads to (neglecting the constant factor
$j_{1}!j_{2}!$)
\begin{displaymath}
       {\bf \Omega}_{n}\vert _{u_1,u_2}\cdot {\bf \Omega}_{0}\vert _{u_1,u_2}
       = \left(\sum_{k}(-)^{k} {n \choose k}
       {j_{1}-n \choose u_{1}-(n-k)} {j_{2}-n \choose u_{2}-k} \right)^{|2|}.
\end{displaymath}
The r.h.s is identical with the square of the components of
${\bf\tilde\Omega}_{n}$ in (\ref{def_tildeomega}). Hence
${\bf\Omega}_{n}\hat ={\bf\tilde\Omega}_{n}^{2}{\bf\Omega}_{0}^{-1}$
has been derived from the Racah formula.


\newpage
\narrowtext


\begin{table}
\begin{tabular}{c|ccccc}
    & \multicolumn{5}{c}{$M=$}\\
	& $-1$ & $-\frac{1}{2}$ & $0$ & $\frac{1}{2}$ & $1$ \\
    \tableline
    $J=\frac{1}{2}$ &  &
    \begin{picture}(10,14) \put(0,10){\vector(1,-1){10}} \end{picture}
    &  &
    \begin{picture}(10,14) \put(0,0){\vector(1,1){10}} \end{picture}
    &   \\
	$J=1$ &
        \begin{picture}(20,22) \put(0,20){\vector(1,-1){10}}
        \put(10,10){\vector(1,-1){10}} \end{picture}
	&  &
	\begin{picture}(20,26)
		\put(0,14){\vector(1,1){10}}
		\put(10,24){\vector(1,-1){10}}
		\put(0,10){\vector(1,-1){10}}
		\put(10,0){\vector(1,1){10}}
	\end{picture}
	&  &
        \begin{picture}(20,22) \put(0,0){\vector(1,1){10}}
        \put(10,10){\vector(1,1){10}} \end{picture}
 	\\
	\vdots &  &  &  &  &
\end{tabular}
\caption{\label{pasc1} Imaginary states of a particle described by the state
$\vert J,M \rangle$}
\end{table}

\begin{table}
\begin{tabular}{c|ccccccccc}
    & \multicolumn{9}{c}{$M=$}\\
        $s(J,M)$ & $-2$ & $-\frac{3}{2}$ & $-1$ & $-\frac{1}{2}$ & $0$ &
        $\ \frac{1}{2}$ &
	$\ 1$ & $\ \frac{3}{2}$ & $\ 2$ \\
	\tableline
	  &  &  &  &  &  &  &  &  & \\
	$J=\frac{1}{2}$ &  &  &  & $1$ &  & $1$ &  &  & \\
	$J=1$ &  &  & $1$ &  & $2$ &  & $1$  & \\
	$J=\frac{3}{2}$ &  & $1$ &  & $3$ &  & $3$ &  & $1$  & \\
	$J=2$ & $1$ &  & $4$ &  & $6$ &  & $4$ &  & $1$   \\
\vdots &  &  &  &  &  &  &  &
\end{tabular}
\caption{\label{pasc2} Number of the imaginary states of a particle described
by the state $\vert J,M\rangle$ }
\end{table}

\begin{table}
		\begin{tabular}{r|r|rrr}
			 & $M_{2}$ & $-1$ & $0$ & $1$  \\
			\tableline
			$M_{1}$ & $u_{1} \backslash u_{2}$ & 0 & 1 & 2  \\
			\tableline
			$-1$ & $0$ & $1$ & $2$ & $1$  \\
			$0$ & $1$ & $2$ & $4$ & $2$  \\
			$1$ & $2$ & $1$ & $2$ & $1$  \\
		\end{tabular}
        \caption{\label{oneone} $M$ and $u$-values of ${\bf\Omega}_{0}$}
\end{table}


\begin{references}
	\bibitem{saku}
        J.J. Sakurai, {\it Modern Quantum Mechanics} (Addison-Wesley,
        Reading, 1994).

	\bibitem{mess}
        A. Messiah, {\it Quantum mechanics, Volume II} (North Holland,
        Amsterdam, 1963).

        \bibitem{cgtables}
        Particle data group, Phys. Rev. D {\bf 50}, 1287 (1994).

	\bibitem{schw}
        J. Schwinger, in: {\it Quantum Theory of Angular Momentum},
        edited by L.C. Biedenharn and H. Van Dam (Academic Press,
        New York, 1965), p.229.

	\bibitem{bronst}
        M. Abramowitz and I.A. Stegun, {\it Handbook of Mathematical
        Functions} (National Bureau of Standards, Washington, 1964).

\end{references}
\end{document}